\documentclass[oldversion, referee]{aa}
\usepackage{times,graphicx,amssymb,lscape}
\usepackage{rotating}
\usepackage{natbib}
\bibpunct{(}{)}{;}{a}{}{,}  

\begin{document}

\title{Opacity in the upper atmospheres of active stars II. AD Leonis}

\author{D.J.~Christian\inst{1}
       \and
	M.~Mathioudakis\inst{1}
        \and
        D.S.~Bloomfield\inst{1}
        \and
        J. Dupuis\inst{2}
        \and
        F.P.~Keenan\inst{1} 
        \and
        D.L.~Pollacco\inst{1}
        \and
        R.F.~Malina\inst{3}
        }

\institute{Department of Pure and Applied Physics, The Queen's University Belfast,
	Belfast, BT7~1NN, Northern Ireland, U.K. 
\and
Department of Physics and Astronomy, Johns Hopkins University,
 3400 North Charles Street, Baltimore, MD 21218.   
\and
Laboratoire d'Astrophysique de Marseille, BP 8, Traverse du Siphon, 13376 Marseille Cedex 12, France}

\offprints{D.J. Christian, \email{d.christian@qub.ac.uk}}

\date{Received date / Accepted date}

\abstract{
We present FUV and UV spectroscopic observations of AD~Leonis, 
with the aim of investigating opacity effects 
in the transition regions of late--type stars.
The \ion{C}{iii} lines in {\sl FUSE} spectra show significant opacity during 
both the quiescent and flaring states
of AD Leonis, with up to 30\% of the expected flux being 
lost during the latter. 
Other {\sl FUSE} emission lines tested for opacity include those of \ion{O}{vi}, 
while \ion{C}{iv}, \ion{Si}{iv} and \ion{N}{v} transitions observed with {\sc STIS} are
also investigated.
These lines only reveal modest amounts of opacity
with losses during flaring of up to 20\%.
Optical depths have been calculated  
for homogeneous and inhomogeneous geometries, 
giving path lengths of $\approx$ 20--60 km and $\approx$ 10--30 km, 
respectively, under quiescent conditions. 
However path lengths derived during flaring are $\approx$2--3 times larger. 
These values are in excellent agreement 
with both estimates of the small-scale structure observed in the solar 
transition region, and path lengths derived 
previously for several other active late-type stars.
}

\authorrunning{D.J. Christian et~al.}
\titlerunning{FUV upper atmosphere opacity on AD Leonis}

\maketitle

\section{Introduction}
 The transition region (TR) in late-type stars shows a temperature
rise from the chromospheric value of a few $\times10^{4}$ K 
to coronal temperatures
of $\approx10^{7}$ K. In the solar case, the TR scale height has
been constrained to $\leq$50 km \citep{DM80, KK86, dos04}. However such
important information on the TR parameters are lacking for most stars.
Several studies have investigated the TR scale heights of active stars, 
finding values of 10 to 100 km for active stars such as AU Mic
(Bloomfield et al.~2002, Paper~I), and the flare star Proxima Centauri \citep{CM04}. 

Previous studies seeking to derive scale height for stellar TRs
\citep{math99, BMC02, CM04} used the relation of optical depth 
to physical parameters, such as electron density and path length, and 
such a method is employed here. 
The optical depth at line center is given by:

\begin{equation}
\label{tau}
\tau_0 = 1.16\times10^{-14} \lambda f_{ij} \sqrt{\frac{M}{T}} \frac{n_{\mathrm{i}}}{n_{\mathrm{el}}}
\frac{n_{\mathrm{el}}}{n_{\mathrm{H}}} \frac{n_{\mathrm{H}}}{n_{\mathrm{e}}} n_{\mathrm{e}} l
\end{equation}
where $\lambda$ is the wavelength in \AA, $f_{ij}$ the oscillator strength of the transition, 
$M$ the mass of the absorbing atom in atomic mass units, $T$ the temperature in K, $l$ the path length in cm and
$n_{\mathrm{i}}$, $n_{\mathrm{el}}$, $n_{\mathrm{H}}$, $n_{\mathrm{e}}$ the number densities
in cm$^{-3}$ of ions in the lower level $i$, element, hydrogen and free electrons, 
respectively \citep{mitc61, J67}. Atomic parameters for the transitions 
of the ions used in this work have been summarized in \citet{CM04}. 

The optical depths for particular emission lines are derived from
ratios of line fluxes that are expected to have certain values in 
optically thin conditions, an assumed geometry, and the escape probability
calculated for that geometry.
For example, the \ion{C}{iii} 1175.71~\AA\ and 1174.94~\AA\ emission lines 
arise from a common upper 
level, and their expected line flux ratio is related to the ratio of their
Einstein A-values. Under optically thin conditions, the ratio of 
the flux of the $\lambda\lambda$1175.71 line to that of 
$\lambda\lambda$1174.94 is expected to be 3.  
Other transitions, such as those of \ion{O}{vi}, \ion{N}{v}, \ion{Si}{iv} and 
\ion{C}{iv}, have a common lower level and under optically thin conditions their 
expected line flux ratio scales as the ratio of their 
electron impact excitation rates, which for those lines 
is expected to be 2. For a detailed description of the method used 
the reader is referred to \citet{BMC02}.

In this paper we continue our in-depth study of opacity effects in stellar TR
with an analysis of the well-known flare star AD Leonis 
(hereafter AD~Leo, dM3.5e; BD+20$\degr$2465, GJ~388). 
The high levels of activity and flaring behavior in AD~Leo have made it a
well studied candidate for flare-activity relations and it has been
the subject of several multi-wavelength campaigns \citep{Ha03}.
Emission lines arising from the transition region are observed to 
show significant downflows, 
in agreement with chromospheric condensation models applied to 
solar flares \citep{Ha03}. Studies of the coronal emission of AD~Leo at X-ray
wavelengths have found compact loop structures of size $\approx$30\% of the star's 
radius \citep{FMR99, S99}.

The {\sl FUSE} and {\sl HST} observations and analysis of AD~Leo are presented in \S\ 2. 
We derive light curves in two different emission lines
from the {\sl FUSE} observations and note AD~Leo was active with several
flares. The temporal information is employed to selected spectra as 
a function of intensity, and we present the results of the spectral
analysis in \S\ 3.
From these spectra we derive ratios of line fluxes for   
emission features of interest, and  
these line ratios are used in \S\ 3.2 to calculate optical depths.
The optical depths, combined with the values of the electron density, allow us 
to calculate the extent of the scattering regions, and we compare these results to 
those from other active late-type stars and the Sun in \S\ 4. 
Lastly, in \S\ 5, we present
 the conclusions of our findings and compare the results 
to the solar case and previous work on AU~Mic, YZ~CMi and Proxima Centauri.

\medskip

\section{Observations \& Analysis}
\label{obs_analy}
{\sl FUSE} observed AD~Leo starting at 03:01 UT 2001 April 11 as part of program A022.
The observations were 
made using the low resolution aperture (LWRS) in time-tagged mode and
covered 41 satellite orbits for a total on source exposure time of 65 ks. 
A complete review of the {\sl FUSE} telescope and instruments can be found in 
\citet{moos00} and \citet{sah00a, sah00b}.
Briefly, {\sl FUSE} consists of four co--aligned prime--focus telescopes, with 
two telescopes having SiC coatings and optimized for the 905--1105 \AA~region and the  
others using LiF coatings to cover 987--1187 \AA. These channels
cover important emission lines for our TR study, such as 
\ion{C}{iii} 977 and 1176 \AA,
and \ion{O}{vi} 1032 and 1038 \AA.

FUSE data were reduced with the latest available version (3.1.0) of the \emph{calfuse} pipeline. 
The results were verified using the previous version
of the pipeline (v2.4) and found in agreement.
Spectra were extracted from the large aperture (LWRS) and these were background
subtracted, flat--fielded, wavelength and flux calibrated as described in the 
\emph{calfuse}
reference guide \citep{dix01}. 
We have combined spectra from different detector channels to increase 
the signal to noise ratio. This was accomplished using the FUSE\_REGISTER\footnotemark{}  
tool provided by the {\sl FUSE} project. 
\footnotetext{http://fuse.pha.jhu.edu/analysis/fuse\_idl\_tools.html} 
Combined spectra for the wavelength regions of 
interest are: segments 1BSiC and 2ASiC for \ion{C}{iii} 977 \AA,
segments 1ALiF and 1ASiC for \ion{O}{vi} 1032 \AA\ and \ion{O}{vi} 1038 \AA, and  
segments 2ALiF and 1BLiF for the \ion{C}{iii} 1176 \AA\ multiplet.
All spectra were binned to a 4--pixel resolution which corresponds to 
$\sim 8$~km s$^{-1}$, approximately half of a formal resolution element. 

\begin{figure}
\includegraphics[scale=0.50,angle=90]{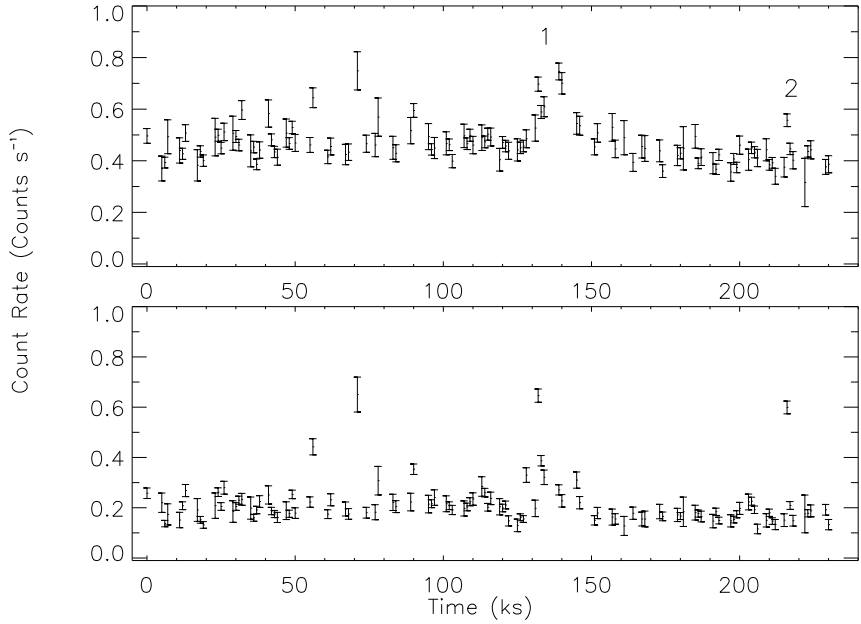}
\caption{AD Leo light curves for the entire {\sl FUSE} observation.
The top panel shows the \ion{O}{vi} 1032 \AA\ light curve and 
bottom panel shows the \ion{C}{iii} 1176 \AA\ multiplet light curve. 
The start time for the first bin in MJD 52010.12598 and
bin size is 1000 s for both light curves.
Flares ``1", and ``2", are noted in the top panel. 
}
\label{lc1}
\end{figure}

Timing analysis of {\sl FUSE} observations of AD~Leo revealed a high level
of activity with several flares. 
Light curves for the entire observations extracted in the lines 
of \ion{O}{vi} 1032 \AA\ 
and the total \ion{C}{iii} 1176 \AA\ multiplet are shown in Figure~\ref{lc1}. 
We concentrated on the two largest flares with both complete
rise and decay times (labeled 1 and 2), and extracted spectra from
these flare intervals. 
Flare 1 had rise times $\leq$300 s for both \ion{O}{vi} 1032 \AA\
and \ion{C}{iii} 1176 \AA\ multiplet light curves. However, the
e-folding decay times for the flare were much shorter for 
\ion{C}{iii}, being $\approx$100~s while the decay time was
nearly 1000~s for \ion{O}{vi}.
Similarly the rise times for flare 2 were $\approx$200~s for both
ions, and the flare decay time was $\approx$ 30~s for 
\ion{C}{iii} and nearly 10 times longer for \ion{O}{vi}.
FUV flares observed in \ion{C}{iii} were also found to decay more
rapidly for Prox Cen \citep{CM04}.
The luminosities of flares 1 and 2 are  
$\approx$8 $\times 10^{26}$~erg s$^{-1}$ for the total flare  
in the \ion{C}{iii} 1176~\AA\ line and about 5 times higher for the
flare peaks. These are about an order of magnitude higher than those
observed for Prox Cen \citep{CM04}.  The total flare 1 and 2 luminosities are 
9.3 $\times 10^{25}$~erg s$^{-1}$ and  
7.1$\times10^{26}$~erg s$^{-1}$ in the \ion{O}{vi} 1032 \AA\ transition, 
respectively. The \ion{O}{vi} 1032 \AA\ peak flare luminosities
are approximately 50\% higher.

We have also obtained archival Hubble Space Telescope 
Imaging Spectrometer ({\sc STIS}) data to investigate important FUV lines
which may also show opacity effects, such as 
\ion{C}{iv}, \ion{Si}{iv} and \ion{N}{v}, as well as further measurements
of \ion{C}{iii}.
Additionally, \ion{O}{iv} lines were used to constrain the
electron densities.
AD~Leo was observed as part of a multi-wavelength campaign under
{\sl HST} program 8613 (root O61S; PI Hawley) on 2000 March 3.  
These observations were performed with the E140M grating 
covering 1150 -- 1700 \AA,  0.2$\arcsec\times0.2\arcsec$ aperture, 
and FUV-MAMA detector, yielding a resolution of $\approx$70,000, and 
providing us with both quiescent and flaring spectra \citep{Ha03}.
We also obtained the {\sc STIS} observations of AD Leo on
2002 June 1 (program 9271; O6JG; PI A. Brown), which  
provided an additional quiescent spectrum.
The 1-D extracted spectra were reduced and combined with the standard
HST\_CALIB.STIS IRAF routines. The O6JG found AD~Leo in a 
quiescent state with an average flux of 
1.67$\times10^{-13}$~erg~cm$^{-2}$s$^{-1}$ 
(luminosity = 4.4$\times10^{26}$~erg~s$^{-1}$) 
in the \ion{C}{iii} multiplet. 
However the earlier
O61S observations found AD~Leo to be very active.  We created a
total flare spectrum with a \ion{C}{iii} multiplet flux of 
3.26$\times10^{-12}$~erg~cm$^{-2}$~s$^{-1}$ and a spectrum from the flare peak
with a flux of 5.14$\times10^{-12}$~erg~cm$^{-2}$~s$^{-1}$. 
Our derived fluxes for the quiescent level of AD~Leo were about
10\% lower than those used by \citet{Ha03} for the O61S observation, and this
reflects slightly different interval selections. The same holds for
the largest AD~Leo flare observed with {\sc STIS} (Hawley's flare 8). Our selection
of a larger flare interval has $\approx$50\% less flux than the Hawley peak 
spectrum, but has similar line flux ratios.

As in our previous work \citep{BMC02, CM04}, we derived 
line fluxes by fitting Gaussians to the observed line profiles. 
In general, a 
single Gaussian profile was sufficient to fit the \ion{O}{vi} and \ion{C}{iii} 
transitions. However, a small red component was detected in the profiles 
of some of the {\sc STIS}
\ion{N}{v}, \ion{Si}{iv}, and \ion{C}{iv} lines
 and two Gaussians were fitted to these in both the quiescent and flaring
spectra.
The six component \ion{C}{iii} 1176 \AA\ multiplet was 
fitted with five separate Gaussians, 
with the 1175.59 \AA\ and 1175.71 \AA\ lines being treated as a single feature,
and the central wavelength of each Gaussian was fixed relative to the
1174.9 \AA\ line (at the laboratory wavelength separation value),
but the wavelength of the latter was left as a free parameter. In this way
any shifts caused by the spectrometers could be taken into account. The 
Gaussian profile widths were set to be equal for all lines, but the 
actual values were left as free parameters. 

\begin{figure}
\includegraphics[scale=0.50,angle=90]{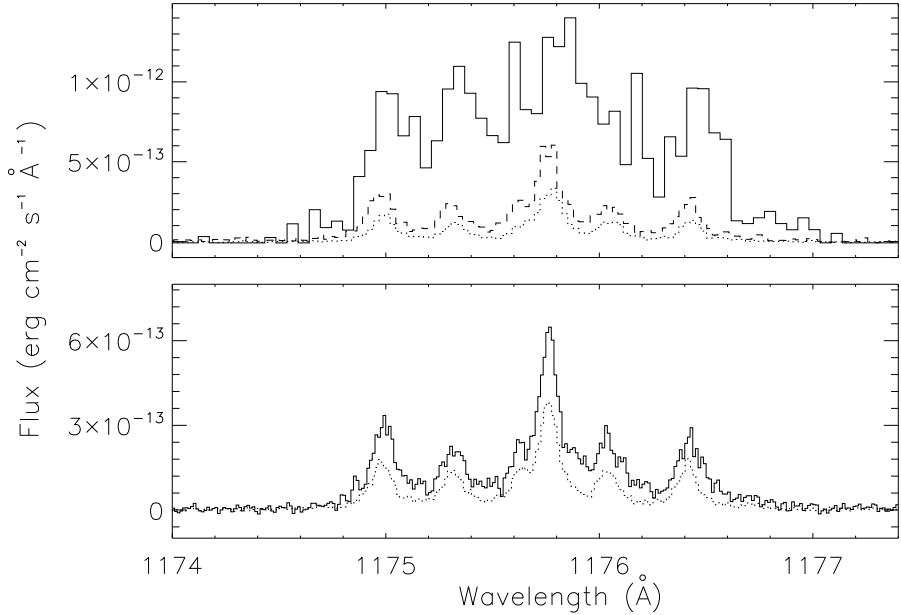}
\caption{ {\sc STIS} and {\sl FUSE} \ion{C}{iii} 1176 \AA\ multiplet spectra. The top panel
shows the {\sl FUSE} spectra: including the spectrum from the flare peak
(solid line), total flare (dashed line), and quiescence (dotted line).
The lower panel shows the {\sc STIS} spectra for the total flare (solid line)
and quiescence (dotted line). (Note these spectra were not obtained 
simultaneously). 
}
\label{CIII_1175_total_spec}
\end{figure}

\section{Results}
\subsection{Opacity derived using Line Ratios}
The ratio of the \ion{C}{iii} 1175.7 to 1174.9 \AA\ 
line fluxes is proportional to the ratio of their Einstein A values, and
is expected to be 3 under optically thin conditions. 
SUMER observations have shown this ratio approaches the optically thin value 
near the Sun's center, but deviates significantly for path length approaching the solar limb
\citep{B00}. The {\sc SUMER} results are consistent with the earlier results of 
\citet{DM80} using {\sl Skylab} data. 
For AD~Leo, the observed \ion{C}{iii} ratios show deviations from the 
optically thin approximation in all spectra, both flaring and quiescent.
We show the \ion{C}{iii} 1176 \AA~ spectra for 
the {\sl FUSE} quiescent, total flare 1 and flare 1 peak spectra and
the {\sc STIS} quiescent and flare spectra 
in Figure~\ref{CIII_1175_total_spec}.
The quiescent spectrum has a \ion{C}{iii} ratio of 2.24$\pm$0.17,
while the spectra extracted from the flare peaks show the largest deviations 
from the optically thin case, with ratios of 1.47$\pm$0.16 and 1.19$\pm$0.11 
for flare 1 and 2 peaks, respectively. 
In Table~\ref{tab1} the \ion{C}{iii} and \ion{O}{vi} fluxes and line
ratios are presented.
\begin{table*}
\begin{center}
\caption{Line fluxes and ratios of the \ion{C}{iii} multiplet and \ion{O}{vi} resonance lines
observed from AD Leo with {\sl FUSE} and {\sc STIS}. 
The fluxes are given in units of 10$^{-14}$ erg cm$^{-2}$ s$^{-1}$
followed by the estimated error in brackets. The fluxes given for each of the flares
are the total observed fluxes minus the quiescent components.}
\begin{tabular}{llllllllllllll}
\hline
\noalign{\smallskip}
\tiny
        & \multicolumn{6}{c}{\ion{C}{iii}} & & \multicolumn{6}{c}{\ion{O}{vi}} \\
\cline{2-7} \cline{9-13}
        & \multicolumn{2}{c}{Flux} &  &  &
 \multicolumn{2}{c}{$\tau_{1175.7}$} & & \multicolumn{2}{c}{Flux} &
            &  &  \multicolumn{2}{c}{$\tau_{1031.9}$}\\
\cline{2-3} \cline{6-7} \cline{9-10} \cline{13-14}
            &  1175.7 \AA &  1174.9 \AA & Ratio    & 
 $\frac{P_{\mathrm{thick}}}{P_{\mathrm{thin}}}$  & Hom.  &  Inh. &   & 1031.9 \AA  &  1037.6 \AA &  Ratio  &  $\frac{P_{\mathrm{thick}}}{P_{\mathrm{thin}}}$       &    Hom.    &  Inh.
 \\ 
\noalign{\smallskip}
\hline
\noalign{\smallskip}
Quiet
    &6.01(0.20) & 2.68(0.18) &2.24(0.17) &0.74  & 1.55 & 0.67  &
    &26.83(0.20)& 13.75(0.19)& 1.95(0.03)& 0.97 & 0.18 & 0.08          \\
F1TOT$^{a}$
    &5.69(0.26) & 3.30(0.25) &1.72(0.15) &0.57  & 3.82 & 1.32 &
    &8.86(0.32) & 5.29(0.72) &1.67(0.24) &0.84  & 1.13 & 0.50     \\   
F1PK
   &35.37(2.17) &24.04(2.23)& 1.47(0.16) &0.49  & 6.39& 1.76 &
   & 20.80(2.11)& 14.06(1.62)& 1.48(0.23)& 0.74 & 2.32& 0.91       \\
F2TOT
  & 4.33(0.24) & 3.17(0.24) &1.37(0.13) &0.45  & 9.21 & 2.02 &
  & 1.43(0.23) & 0.90(0.29) &1.59(0.57) &0.79 & 1.65 & 0.69 \\ 
F2PK
  & 28.41(1.68) &23.91(1.66) &1.19(0.11)& 0.39  &14.5 & 2.52 &
  & 20.30(2.05) & 13.61(1.58)& 1.49(0.23)& 0.75 & 2.17 & 0.86   \\ 

\\
{\sc STIS}\\
Quiet$^{b}$
       & 7.00(0.19) &2.72(0.18)  &2.57(0.18)   &0.85  & 0.76 & 0.35   &
       &            &            &             &      &      &       \\
Quiet$^{c}$
       &6.35(0.19)   &2.71(0.17)  &2.34(0.16)  &0.65  & 2.49 & 0.98   &
       &            &            &             &      &      &       \\
Flare 
       &5.10(0.18)  &3.07(0.17)  &1.65(0.11)   &0.55  & 4.29 & 1.42   &
       &            &            &             &      &      &       \\
Fl PK 
       &9.46(0.49)  &5.80(0.48)  &1.63(0.16)   &0.54  & 4.55 & 1.47   &
       &            &            &             &      &      &       \\
\noalign{\smallskip}
\hline
\label{tab1}
\end{tabular}
\end{center}
\footnotesize
$^{a}$: {Flare Total and Peak emission abbreviated as TOT and PK, respectively.} \\
$^{b}$: {{\sc STIS} O6JG} \\
$^{c}$: {{\sc STIS} O61S} 
\end{table*}

The ratio of the total emission of the \ion{C}{iii} 1176 \AA~multiplet 
to that of the \ion{C}{iii} 977 \AA\ line has a density dependence in 
the $10^8 - 5 \times 10^{10}$~cm$^{-3}$ range. 
We calculated a \ion{C}{iii} $\lambda1176/\lambda977$ ratio of 1.3$\pm$0.10
for the quiescent spectrum and $>$ 1.5 for all flaring spectra.
Using CHIANTI \citep{dere97} 
to calculate theoretical line ratios indicates 
$n_e > 10^{11}$ cm$^{-3}$ for the quiescent, total flare
and flare peak spectra.
However, as pointed out in previous work \citep{CM04}, 
the \ion{C}{iii} 977 \AA\ emission line is optically 
thick, and any diagnostics that use this line provide 
only a lower limit to the density. We make use of other diagnostic lines 
to derive the electron density in the next section.    
However, we can estimate the amount of flux lost from the 
1176 \AA\ multiplet 
by scaling from the 1174.93 \AA\ line, which we 
assume to be optically thin. 
Under optically thin conditions the 1174.93 \AA\ line contributes
about 14\% of the total 1176 \AA\ multiplet
flux, and as the optical depth increase its
relative contribution increases.
We find that 80\% of the expected \ion{C}{iii} 1176 flux 
is observed during quiescence, 
while 70\% and 80\% of the expected flux is observed for 
flares 1 and 2, respectively. 

The {\sl FUSE} \ion{O}{vi} 1032 and 1038 \AA~resonance lines and {\sc STIS}
\ion{N}{v}, \ion{Si}{iv} and \ion{C}{iv} 
resonance lines also provide constraints on the transition region
scale heights. 
For the optically thin case, the ratio of 
their line fluxes should be 2. 
There is no strong  evidence for significant opacity in the quiescent spectrum
with an observed \ion{O}{vi} ratio of 1.95$\pm$0.03.
However, the flares show a modest amount of opacity with the flare 1 and 
flare 2 peak spectra showing similar ratios of 1.48$\pm$0.23.
Similarly, the observed \ion{N}{v} quiescent spectrum 
shows no deviation from the optically thin case with a ratio of
2.20$\pm$0.06, and the \ion{N}{v} flare and flare peak
spectra are optically thin within the derived uncertainties.  
Deviations from optically thin conditions 
are detected for the {\sc STIS} flare spectra for \ion{Si}{iv}
and \ion{C}{iv} in the quiescent spectra and 
flaring spectra. The \ion{Si}{iv}
and \ion{C}{iv} flare peak spectra show the greatest opacity, with
ratios of 1.49$\pm$0.21 and 1.61$\pm$0.18, respectively.    
{\sc STIS} line fluxes and derived ratios are summarized for the quiescent and 
flaring spectra in Table~\ref{tab2}. 
Comparison of the {\sl FUSE} and {\sc STIS} spectra of the 
resonance lines is presented in the next section. 

\begin{table*}
\begin{center}
\caption{Line fluxes and ratios of the \ion{C}{iv}, \ion{Si}{iv}, 
and \ion{N}{v} doublets observed with STIS.
The fluxes are given in units of 10$^{-14}$ erg cm$^{-2}$ s$^{-1}$
followed by the estimated error in brackets. The fluxes given for each 
of the flares are the total observed fluxes minus the quiescent components.
}
\label{tab2}
\begin{tabular}{lllllll}
\hline
\noalign{\smallskip} 
Ion  & \multicolumn{2}{c}{Flux} & Ratio &
$\frac{P_{\mathrm{thick}}}{P_{\mathrm{thin}}}$ & \multicolumn{2}{c}{$\tau_{ion}$}  \\ 
\cline{6-7}
&      &    &     &     & Hom.    &  Inh.   \\      

\noalign{\smallskip}
\hline
\noalign{\smallskip}

\ion{N}{v}   & 1238.7   & 1242.8      &           &      & \multicolumn{2}{c}{$\tau_{1238.7}$}       \\
O61S Quiet   &9.0(0.14)  &4.1(0.1)    & 2.20(0.06) &1.10 & ... & ...    \\
Flare  & 2.22(0.14) & 1.24(0.13)& 1.79(0.21) &0.89 &0.71 & 0.33   \\
Flare Peak & 4.05(0.62) & 2.19(0.37)& 1.85(0.42) &0.92 &0.49 & 0.23   \\
O6JG Quiet &8.8(0.17)   & 4.1(0.52) &2.14(0.26)  &1.07 & ... & ...    \\
\\
\ion{Si}{iv}  & 1393.8     & 1402.8    &           &      & \multicolumn{2}{c}{$\tau_{1393.8}$}   \\
O61S Quiet   & 8.73(0.19) &4.86(0.10) &1.79(0.05)  &0.90 &0.63 &0.30   \\
Flare       & 8.58(0.19) & 5.57(0.48)& 1.54(0.14) &0.77 &1.90 &0.77   \\
Flare Peak  & 16.15(0.69)& 10.84(1.46)& 1.49(0.21)&0.75 &2.17 &0.86   \\
O6JG Quiet  &8.9(0.18)   &4.70(0.15)  &1.89(0.07) &0.95 &0.29 &0.14 \\
\\
\ion{C}{iv}   & 1548.2   & 1550.8      &           &      & \multicolumn{2}{c}{$\tau_{1548.2}$}       \\
O61S Quiet   &30.1(0.4)   &17.8(0.3)   &1.69(0.04) &0.84 &1.10&0.50   \\
Flare     & 22.41(0.79)& 13.42(0.49)& 1.67(0.09)&0.83 &1.23&0.54   \\
Flare Peak   & 43.84(2.42)& 27.24(2.73)& 1.61(0.18)&0.80 &1.54&0.65   \\
O6JG Quiet   &30.0(0.62)  &17.8(0.38)  &1.68(0.05) &0.84 &1.10&0.50   \\
\noalign{\smallskip}
\hline
\end{tabular}
\end{center}
\end{table*}

\subsection{Line Ratio constraints on plasma emission size}
 We can now use the derived line flux ratios to estimate the 
escape probabilities for our lines of study, and from these 
and an assumed geometry we can then estimate each line's  
optical depth. 
For pairs of lines arises from a common upper level 
(\ion{C}{iii} 1175.71 \AA\ and 1174.94 \AA\ in the present study), 
the ratio
of their fluxes is the ratio of their Einstein A-values reduced
by the ratio of photon escape probabilities (see eq. 2 in \citet{BMC02}).
Pairs of lines arising in a common lower level (\ion{O}{vi}, \ion{N}{v}, \ion{Si}{iv} and
\ion{C}{iv} in the present study), the ratio of their
fluxes scales as the ratio of their oscillator strengths reduced by
the photon escape probability (see eq. 4 in \citet{BMC02}).
For each of these two categories of line pairs,
we consider both 
homogeneous and inhomogeneous distributions of emitters and absorbers
\citep{kast90, BMC02}. The emitters and 
absorbers are spatially distinct in the inhomogeneous case. 
The analytic form of the escape probabilities is given in 
\citet{BMC02} equations 5 \& 6 for the
homogeneous  case and equations 7 \& 8 for the inhomogeneous one.
We present our optical depth for AD~Leo  derived from each of these distributions in
Table~\ref{tab1}, and note that the values from the 
homogeneous approximation are a factor of 2--3 times larger than
those from the inhomogeneous distribution. 
We now use these optical depths to constrain the extent of the scattering 
region for both the homogeneous and inhomogeneous distributions. 

With estimates of the optical depth 
we can use Equation~1 to
derive  information on the extent of the scattering region 
with the only unknown quantity being the electron density. 
We emphasize that equation (1) has been derived on the assumption of
thermal broadening only. If a non-thermal contribution is
included, the optical depths would decrease and hence the derived 
path lengths would decrease.  There is currently
little information on the non-thermal broadening in stellar transition
regions \citep{Red02} but if solar values are assumed the path
lengths could decrease by up to a factor of 3.
As mentioned above, the \ion{C}{iii} 977 \AA\ emission line is optically 
thick, and does not provide an accurate estimate of density. 
However, a powerful density diagnostic are the \ion{O}{iv} lines near 1400 \AA,
in particular the \ion{O}{iv} $\lambda1399.78/\lambda1404.79$ 
and $\lambda1407.38/\lambda1404.79$ ratios formed around log(T) = 5.1.
These are available from the {\sc STIS} observations in the same echelle order
as \ion{Si}{iv}, and we have derived both \ion{O}{iv} ratios for both {\sc STIS} 
quiescent and the total flare spectra (with a better signal to noise than the
flare peak spectra).  
The \ion{O}{iv} $\lambda1399.78/\lambda1404.79$ ratio (R$_1$) has values of
0.96$\pm$0.11 and 1.20$\pm$0.20 for the quiescent and flaring, respectively.
Similarly, the $\lambda1407.38/\lambda1404.79$ ratio (R$_2$)
is 1.06$\pm$0.13 and 1.15$\pm$0.23 
for the quiescent and flaring, respectively. 
The \ion{O}{iv} R$_1$ and R$_2$ quiescent values imply densities
of 3 $\times 10^{10}$ cm$^{-3}$ and 4.5 $\times 10^{10}$ cm$^{-3}$
\citep{keen02}, and we therefore adopt a quiescent plasma density of
4 $\times 10^{10}$ cm$^{-3}$. 
For the flare, the R$_1$ and R$_2$ ratios both imply densities
of $\approx$6 $\times 10^{10}$ cm$^{-3}$ and we therefore adopt this value.

\begin{figure}
\includegraphics[scale=0.50,angle=90]{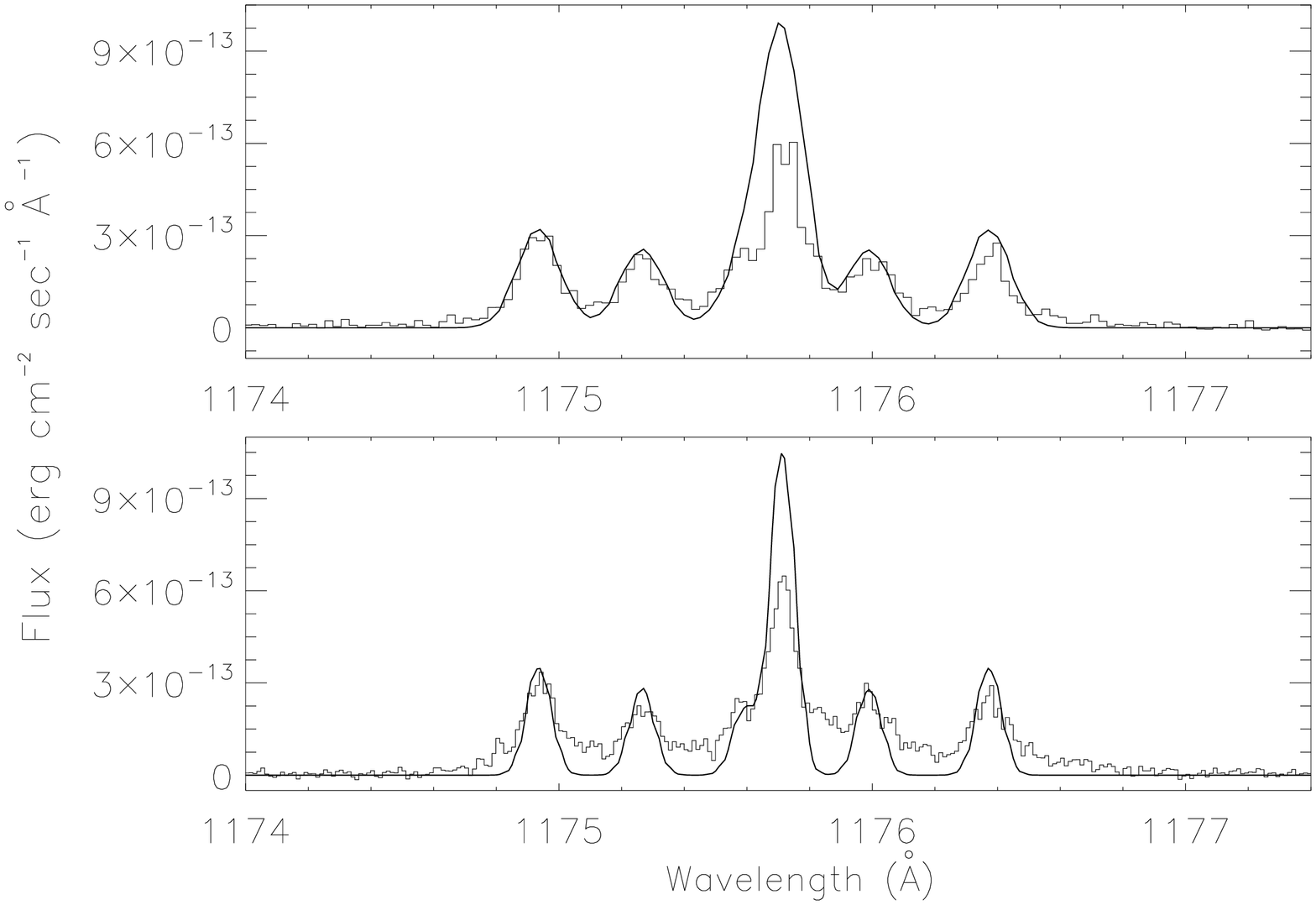}
\caption{Shown are the theoretical line profiles (solid line) over-plotted
on the observed spectra (solid histogram).  The top panel show the {\sl FUSE}
\ion{C}{iii} 1176 \AA\ spectrum during flare 1, and the lower panel show
the {\sc STIS} \ion{C}{iii} spectrum also during  a flare. The line profiles were
generated with CHIANTI using our derived density (6 $\times 10^{10}$ cm$^{-3}$)
 and solar flare DEMs for both panels (please see text).
The theoretical profiles are scaled to match the observed flux from the
optically thin 1174.9 \AA\ line.
}
\label{spec_ciii_chianti}
\end{figure}

\begin{figure}
\includegraphics[scale=0.50,angle=90]{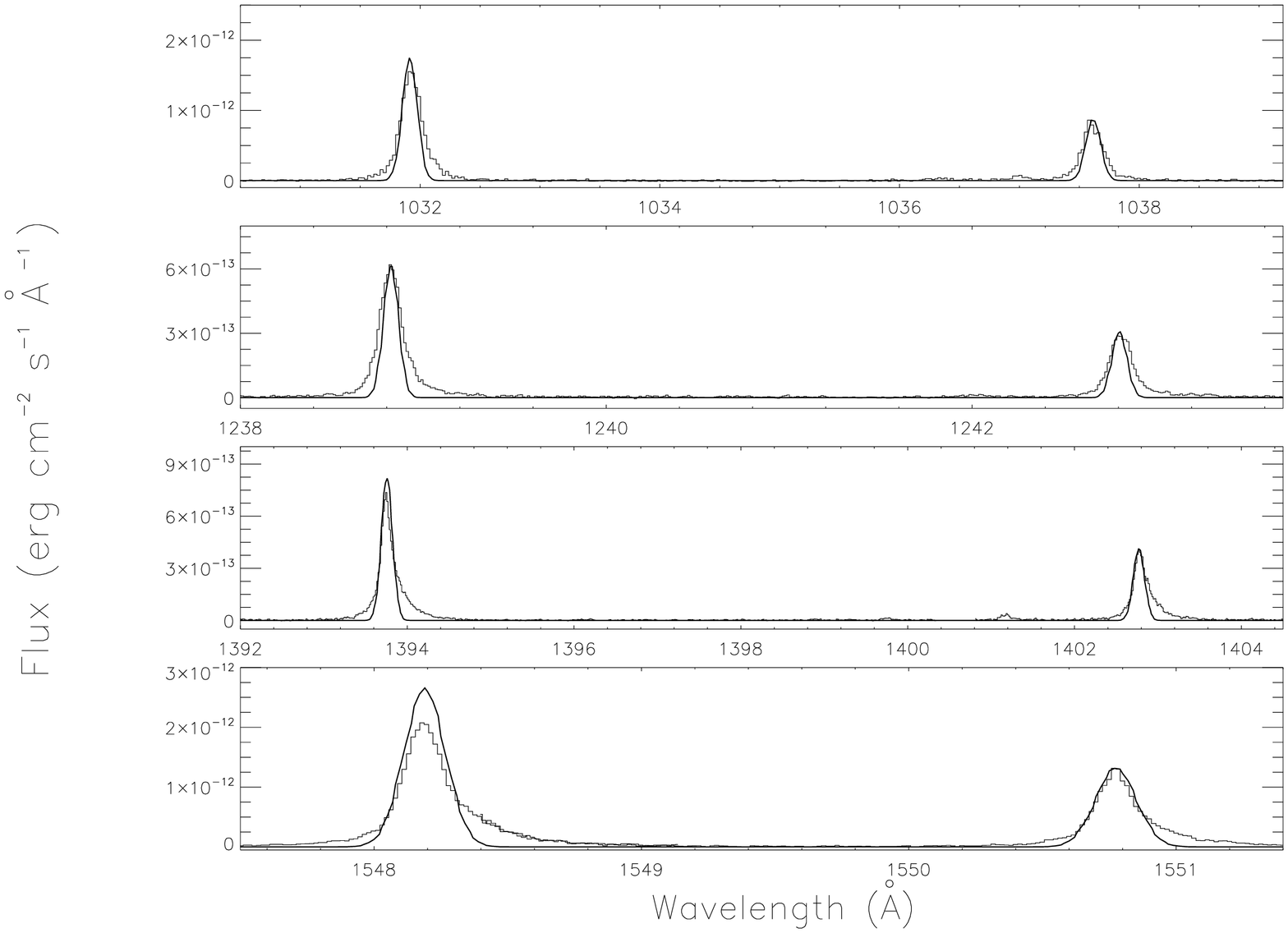}
\caption{Shown are the theoretical line profiles (solid line) over-plotted
on the observed spectra (solid histogram).  The top panel show the {\sl FUSE}  
\ion{O}{vi} spectrum during flare 1, and the lower panels show
the {\sc STIS} NV, Si IV and \ion{C}{iv} spectrum during the flare. 
The line profiles were
generated with CHIANTI using our derived density (6 $\times 10^{10}$ cm$^{-3}$) and solar flare DEMs (please see text).    
As in Fig~3, the theoretical profiles are scaled to match the observed 
flux from the optically thin components, each occurring at longer wavelengths.
}
\label{Chianti_spec}
\end{figure}

We show a comparison of the observed {\sl FUSE} and {\sc STIS}
spectra and the CHIANTI model for the \ion{C}{iii} multiplet in
Figure~\ref{spec_ciii_chianti}. The CHIANTI models were derived with 
the above densities and
for a solar flare differential emission measure (DEM).  
We also show a comparison of the theoretical line profiles
and the observed {\sl FUSE} \ion{O}{vi} and {\sc STIS}
\ion{N}{v}, \ion{Si}{iv}, and \ion{C}{iv} 
spectra in Figure~\ref{Chianti_spec}. Line profiles were generated with
the CHIANTI spectral synthesis package for
the above derived density, and a solar flare DEM for the flare 
spectra.
The opacity effects for the \ion{C}{iii} 1175.71 \AA\  
and \ion{C}{iv} 1548.2 \AA\ lines are clearly visible in the figures. 
The observed profiles also show clear redshifted components. These components
were accounted for in the spectral fitting to derive the line fluxes, 
but are not discussed here.
The redshifted components were discussed in detail for AD~Leo by 
\citet{Ha03}.

Using the electron density estimates from the \ion{O}{iv} line ratios
and the optical depths we can now derive values for the
effective size of the scattering region (Eq 1.) for both quiescent and flare spectra,
and for both homogeneous (Hom) and inhomogeneous (Inh) source geometries.
For the quiescent {\sl FUSE} \ion{C}{iii} spectra, we find path lengths
{\em Hom}$_{\mathrm{CIII}} \approx 36$~km  
and 
{\em Inh}$_{\mathrm{CIII}} \approx 16$~km for homogeneous and
inhomogeneous geometries, respectively. The {\sc STIS} values for the
\ion{C}{iii} path lengths are approximately 50\% larger than those from {\sl FUSE}.
This may be due to AD~Leo being slightly more active in its quiescent state
during the {\sc STIS} observations and possibly also because our {\sc STIS} quiescent
spectrum may be contaminated with several smaller flares. The O6JQ observation
has a path length about 50\% of the derived {\sl FUSE} values.   
For the flaring {\sl FUSE} \ion{C}{iii} spectra, we find
{\em Hom}$_{\mathrm{CIII}} \approx 65$~km  
and {\em Inh}$_{\mathrm{CIII}} \approx 22$~km path lengths for homogeneous and
inhomogeneous geometries for flare 1, respectively, and the spectra extracted
from the flare 1 peak also show similar values. The derived path lengths for
flare 2 are about 50\% larger, but again the flare and flare peak are in 
agreement. 
The {\sc STIS} \ion{C}{iii} flare spectra give very similar
path lengths to those derived for the {\sl FUSE} flare spectra. 
 The path lengths are summarized in Table~\ref{tab3}. 
\begin{table}
\begin{center}
\caption{Table of derived path lengths for observed ions}
\label{tab3}
\begin{tabular}{lll}
Ion   & \multicolumn{2}{c}{Path Length (km)} \\
      & Hom.   &  Inh.   \\
\noalign{\smallskip}                                                            \hline                                                                          \noalign{\smallskip}                                                            
\ion{O}{vi}       &        \\
 ...Quiet     &  24 & 11 \\
 ...Flare 1   & 109 & 48 \\
 ...Flare 1 Peak  & 224 & 88 \\
 ...Flare 2   & 160  & 67    \\
\\
\ion{C}{iii}   \\
 ...Quiet     & 36   & 16        \\
 ...Quiet ({\sc STIS}) & 58    & 23         \\
 ...Flare ({\sc STIS}) & 73 & 24 \\
 ...Flare1    & 65   & 22       \\
 ...Flare 1 peak& 66   & 18     \\
 ...Flare2    & 156  & 34        \\
\\
\ion{C}{iv}  \\
  ...Quiet ({\sc STIS})   & 62    &  28       \\
  ...Flare ({\sc STIS})   & 50    &  22        \\
\\
\ion{Si}{iv} \\
  ...Quiet ({\sc STIS})   & 24    &  11       \\
  ...Flare ({\sc STIS})   & 52    &  21        \\
\\
\noalign{\smallskip}
\hline
\end{tabular}
\end{center}
\end{table}

{\sc STIS} also provided several other resonance line 
opacity measurements, namely \ion{N}{v}, \ion{Si}{iv}, and \ion{C}{iv}. 
The flare spectra for all three ions show strong opacity effects and
deviate from the optically thin approximation. 
The extent of the scattering region derived from the \ion{C}{iv} and
\ion{Si}{iv} flare spectra are similar, with   
 {\em Hom}$_{\mathrm{CIV}} \approx 50$~km  
and {\em Inh}$_{\mathrm{CIV}} \approx 20$~km. 
During quiescence, the \ion{C}{iv} and \ion{Si}{iv} ratios show significant
opacity, but the \ion{N}{v} lines show no opacity effects.
The \ion{C}{iv} path lengths are
{\em Hom}$_{\mathrm{CIV}} \approx 60$~km
and {\em Inh}$_{\mathrm{CIV}} \approx 30$~km,
while \ion{Si}{iv} indicates 
{\em Hom}$_{\mathrm{SiIV}} \approx 24$~km
and {\em Inh}$_{\mathrm{SiIV}} \approx 10$~km.
These values are also summarized in Table~\ref{tab3}.
 
\section{Discussion}
During the flaring activity of AD~Leo observed with {\sl FUSE} 
the optical depths increased by 
factors of 6--9. Although this change in optical depth does increase 
the physical extent 
of the scattering layer, as we adopted a 50\% higher
density for the flare spectra, the flare scattering layers have
similar sizes  to those derived from the quiescent spectra. 
We compare the path lengths from our current results for AD~Leo to those
of the Sun, AU~Mic \citep{BMC02}, YZ~CMi \citep{math99}, and Prox Cen \citep{CM04} in Table~\ref{tab4}.  
We find the quiescent path lengths for OVI are about a factor of 2 larger 
for AD~Leo than AU~Mic. The AD~Leo quiescent path lengths for C~III 
are approximately twice those of AU~Mic and a factor of three larger than 
those of Prox Cen. Similarly the
\ion{C}{iv} results for AD~Leo in quiescence are a factor of 2 larger than
Prox Cen and three times larger than YZ~CMi for the inhomogeneous case.
AD Leo, with log(L$_X$/L$_{bol}$) of -3, is much more active 
than Prox Cen (log(L$_X$/L$_{bol}$) $\approx$ -3.8) and about 50\% more active
than YZ CMi (L$_X$/L$_{bol}$$\approx$-3.3). AU~Mic has a similar activity
level, but had a \ion{C}{iii} peak flare flux about 3 times smaller than
those observed for AD~Leo, and this higher flux may explain the much 
larger path lengths for the latter. 

\begin{table}
\begin{center}
\caption{Comparison of path lengths}
\label{tab4}
\begin{tabular}{lcccc}
\hline
\noalign{\smallskip}
Star &  \multicolumn{4}{c}{Path Length (km)} \\
     & \multicolumn{2}{c}{Quiet}   &  \multicolumn{2}{c}{Flare}   \\
     & Hom.   &  Inh.  & Hom.   &  Inh.   \\
\noalign{\smallskip}
\hline
\noalign{\smallskip}
\ion{O}{vi} &        \\
 AU Mic$^{a}$   & 10   &  5 & ...  & ...   \\
 Prox Cen$^{b}$ & ... & ... &  380 &  120     \\
 AD~Leo  & 24  &  11 & 100-200  & 48-200  \\
\\
\ion{C}{iii}   \\
 Sun$^{c}$    & 50 & ...   &   ...  & ...        \\ 
 AU Mic   & 20   &  7 &  100 & 20    \\
 Prox Cen & 13  &  6  & 24--80   & 8--25     \\
  AD~Leo  & 36  & 16  & 60---160 & 12--34     \\
\\
\ion{C}{iv}  \\
  YZ CMi$^{d}$   & ... &  5  & ...   & ...     \\
  Prox Cen & 30  & 15  & ...   & ....              \\
  AD Leo   & 62  & 28  & 50    & 22          \\ 
  

\\
\noalign{\smallskip}
\hline
\end{tabular}
\end{center}
$^{a}$: AU~Mic results from \citet{BMC02}  \\
$^{b}$: Prox~Cen results from Christian et~al. (2004)  \\
$^{c}$: Solar results from \citet{DM80}  \\
$^{d}$: YZ~CMi results from \citet{math99} 
\end{table}

\section{Conclusions}
\label{conc}
We study the effects of opacity in the 
transition region of the active, late-type star AD Leonis. 
As shown previously, opacity can be used as a diagnostic tool, 
since the optical depth is 
proportional to the line of sight path length. Hence, optical depths 
combined with independent measurements of the electron density allow
direct estimates of the path lengths in the transition region.  
The \ion{C}{iii} lines show significant opacity during 
both the quiescent and flaring states
of AD Leo, with up to 30\% of the expected flux being 
lost in the latter. Analysis of other FUV resonance lines also showed 
opacity effects, but with $\approx$20\% of the expected flux being lost. 
For AD~Leo discussed here, we find path lengths during quiescence 
in the range of $\approx$ 20--60 km and $\approx$ 10--30 km 
for homogeneous and inhomogeneous geometries, 
respectively. Path lengths derived during flaring were $\approx$2--3 times larger. 
These path lengths are very similar to those found for other
active, late-type stars and that of the
solar transition region \citep{KK86, dos04}. 
Therefore, solar and stellar transition
regions must have similar spatial characteristics, and it may be
fruitful to compare transition region sizes for hotter
stars further up the main-sequence.  

\acknowledgements
This research was partially supported by a NASA {\sl FUSE} Guest-Investigator 
Grant (NNG04GC75G). 
FPK is grateful to AWE Aldermaston for the award of a William Penney Fellowship.
Data sets presented here were obtained from the Multimission Archive
at the Space Telescope Science Institute (MAST).  STScI is operated by the 
Association of Universities for Research in Astronomy, Inc., under NASA 
contract NAS5-26555. Support for MAST for non-HST data is provided by the 
NASA Office of Space Science via grant NAG5-7584 and by other grants and contracts.
\vspace{0.1in}

\bibliography{aa}
\bibliography{submit}
\bibitem[Brooks et al.(2002)]{B00} Brooks, D.~H. et al. 2000, \aap, 357, 697
\bibitem[Bloomfield et al.(2002)]{BMC02} Bloomfield, D.S., Mathioudakis, M., Christian, D.J., Keenan, F.P., \& Linsky, J. 2002, \aap, 390, 219 (Paper~I)
\bibitem[Christian et al.(2004)]{CM04}Christian, D.J., Mathioudakis, M., Bloomfield, D.S., Dupuis, J., Keenan, F P. 2004, \apj, 612, 1140
\bibitem[Dere et al.(1997)]{dere97} Dere, K.P., Landi, E., Mason, H.E., Monsignori--Fossi, B.C., \& Young, P.R. 1997, A\&AS, 125, 149
\bibitem[Doschek et al.(2004)]{dos04} Doschek, G. A. \& Feldman, U. 2004, \apj, 601, 1061 
\bibitem[Doyle \& McWhirter(1980)]{DM80}Doyle, J.G., \& McWhirter, R.W.P. 1980, 
\mnras, 193,947
\bibitem[Favata, Micela, \& Reale(1999)]{FMR99} Favata, F., Micela, G., \& Reale, F. 1999, \aap, 354, 1021
\bibitem[Hawley et al.(2003)]{Ha03}Hawley, S.L. et al, 2003, \apj, 597, 535
\bibitem[Jordan(1967)]{J67}Jordan, C. 1967, \solphys, 2, 441 
\bibitem[Kastner \& Kastner(1990)]{kast90} Kastner, S.O., \& Kastner, R.E. 1990, \jqsrt, 44, 275
\bibitem[Keenan et al.(2002)]{keen02} Keenan, F.P. et al. 2002, \mnras, 337, 901
\bibitem[Keenan \& Kingston(1986)]{KK86}Keenan, F.P., \& Kingston, A. E. 1986, \mnras, 2220, 493 
\bibitem[Mathioudakis et al.(1999)]{math99} Mathioudakis, M., McKenny, J., Keenan, F.P., Williams, D.R., \& Phillips, K.J.H. 1999, \aap, 351, L23
\bibitem[Mitchell \& Zemansky(1961)]{mitc61} Mitchell, A. C. G., \& Zemansky, M. W. 1961, In \emph{"Resonance Radiation and Excited Atoms"}, 
Cambridge University Press
\bibitem[Moos et al.(2000)]{moos00} Moos, H.W., Cash, W.C., Cowie, L.L., et~al. 2000, \apjl, 538, L1 
\bibitem[Redfield et al.(2002)]{Red02} Redfield, S., Linsky, J.L., Ake, T.B, Ayres, T.R., Dupree, A.K., Robinson, R.D., Wood, B.E, \& Young, P.R 2002, \apj, 581, 626 
\bibitem[Sahnow et al.(2000a)]{sah00a} Sahnow, D.J., Moos, H.W., Ake, T.B., et~al. 2000a, \apjl, 538, L7
\bibitem[Sahnow et al.(2000b)]{sah00b} Sahnow, D.J., Moos, H.W., Ake, T.B., et~al. 2000b, In \emph{"UV, Optical, and IR Space 
Telescopes and Instruments" Proc. SPIE}, eds. Breckinridge, J.B., Jakobsen, P., Vol. 4013, 334
\bibitem[Sciortino et al.(1999)]{S99}Sciortino, S., Maggio, A., Favata, F., \& Orlando, S. 1999, \aap, 342, 502
\bibitem[van Dixon et al.(2001)]{dix01} van Dixon, W., Kruk, J., \& Murphy, E. 2001, Calfuse Reference Guide, http://fuse.pha.jhu.edu/analysis/calfuse.html 


\end{document}